# Magic thickness of 25 Å makes periodic metal-insulator transitions

M. Sakoda[1,*], H. Nobukane[2,3], S. Shimoda[4], and S. Tanda[1,3]

[1]*Department of Applied Physics, Graduate School of Engineering, Hokkaido University, Sapporo 060-8628, Hokkaido, Japan*

[2]*Department of Physics, Graduate School of Science, Hokkaido University, Sapporo 060-0810, Hokkaido, Japan*

[3]*Center of Education and Research for Topological Science and Technology, Hokkaido University, Sapporo, 060-8628, Hokkaido, Japan*

[4]*Institute for Catalysis, Hokkaido University, Sapporo 001-0021, Hokkaido, Japan*

Novel quantum phenomena, including high-temperature superconductivity, topological properties, and charge/spin density waves, appear in low-dimensional conductive materials. It is possible to artificially create low-dimensional systems by fabricating ultrathin films, quantum wires, or quantum dots with flat interfaces[1-6]. Some experiments have been performed on ultrathin compounds of strongly correlated electron systems[7,8]. However, since it is technically difficult to control multiple elements precisely, most of the properties of artificially fabricated low-dimensional compounds fall into uncharted territory. Here we show that extraordinary metal-insulator transitions that oscillate depending on the scale occur in $CaRuO_3$ films with a thickness of around several unit cells. We grow high-crystalline $CaRuO_3$ ultrathin films, whose surface roughness is controlled at 199 pm, by molecular beam epitaxy. We observe that resistivity oscillates with a 'magic' thickness of 25 Å, which changes by 3 and 9 orders of magnitude at room temperature and at low temperature, respectively. These changes are much larger than quantum size effects. We also confirm the same periodicity with photoelectron spectroscopy by etching the ultrathin film. Considering the large energy, periodicity and anisotropy, we conclude that the oscillating transitions originate from the commensurability of Mott insulation triggered by Peierls instability arising from a dual restriction on the dimensions in wavenumber space and real space. We have shown the possibility of producing new functional materials by controlling film thickness on electron correlated compounds at the picometer level.

In ultrathin films with a thickness of several nanometers, the degree of freedom of electrons is artificially enclosed in two dimensions. For example, quantum size effects appear in single element ultrathin films whose thicknesses are approximately of the order of de Broglie wavelengths[1-6]. It is necessary to develop ultrathin films with clear interfaces into compounds that have unique electronic states. Ruthenium oxides based on the $RuO_6$ octahedra layered structure have a treasure box of attractive physical properties including an anisotropic superconductor $Sr_2RuO_4$[9], $Ca_2RuO_4$, which exhibits the electric field-induced Mott insulator-metal transition[10], and high-$T_c$ superconductivity in thin flakes[11]. In this study, we focus on growing $CaRuO_3$ ultrathin films as the target material for artificially fabricated low-dimensional electronic systems.

$CaRuO_3$ has a $GaFeO_3$-type crystalline structure consisting of a simple orthorhombic lattice (space group: *Pbnm*)[12,13] (Fig. 1a). Pure $CaRuO_3$ single crystals with a high residual resistivity ratio have been reported for bulk and thin films[14,15]. In this paper, we develop high quality $CaRuO_3$ ultrathin films with a stable supply of molecules by using molecular beam epitaxy (MBE)[16,17] (Fig. 1b) to investigate the phenomena that occur in artificially fabricated two-dimensional (2D) electron systems. $CaRuO_3$ is a strongly correlated electron compound with the 2D-pole Fermi surface (FS)[17]. Our purpose is to explore the low-dimensional electronic properties that arise from dual restriction on the dimensions in wavenumber space by using pole FS and in real space by using ultrathin films with clear interfaces. Since the Fermi length of $CaRuO_3$ is as much as several nanometers[18], it is expected that scale-dependent quantum phenomena will appear in the ultrathin films.

When fabricating ultrathin films, it is important to control their thickness with a smaller than nanometer order. Figure 1c shows the reflection high-energy electron diffraction (RHEED) patterns of a $CaRuO_3$ film with a thickness of 8 Å. A sharp streak pattern is observed even in nanometer-thick films, which proves that the epitaxial films grow with a highly crystalline form and are flat with few impurities[15] (Extended Data Fig. 2). The lattice matting between a $NdGaO_3$ substrate and $CaRuO_3$ is as small as 0.5 or 0.8 % in the (110) plane. Furthermore, the oxygens in the $RuO_6$ octahedron are in matching positions. Consequently, the epitaxiality is excellent[19,20]. Growth reproducibility is also good, suggesting that the self-regulation of the elements of MBE is performed efficiently and oversupplied Ca and Ru are re-evaporated. Figure 1d shows the film surface observed with atomic force microscopy (AFM). The average surface roughness of $CaRuO_3$ ultrathin film is 199 pm, which is half the size of a $RuO_6$ conductive layer.

We measure the electrical resistivity on $CaRuO_3$ ultrathin films at each thickness. The film with a conventional thickness of 504 Å has the same resistivity as the reported metal and is accompanied by a magnetic transition and non-Fermi liquid behavior[21-28] (Extended Data Fig. 3b, c). Surprisingly, the electrical resistivity of $CaRuO_3$ ultrathin films changes systematically by orders of magnitude. Figure 2a shows the temperature dependence of the electrical resistivity on

CaRuO$_3$ ultrathin films with thicknesses between 101 and 85 Å. The film with a thickness of 101 Å is metallic, and the electrical resistivity increases exponentially as the film thickness decreases. Film with a thickness of 85 Å exhibits typical insulating behavior with maximum resistivities of 1x10$^3$ and 2x10$^8$ mΩcm at 300 and 4.2 K, respectively. The magnitudes of the electrical resistivity decrease again and return to metallic behavior in the thinner films (Fig. 2b). The magnitudes of the electrical resistivity on the 71 Å thick CaRuO$_3$ ultrathin film are 3x10$^{-1}$ and 7x10$^{-2}$ mΩcm at 300 and 4.2 K, respectively. Despite the slight difference between the sizes of the films with thicknesses of 71 and 85 Å, there is the large change in electrical resistivity of 3 orders of magnitude at room temperature and 9 orders of magnitude or more at low temperature. The dependence of electrical resistivity on scale has not been reported in the thick films or bulk samples on CaRuO$_3$. Figure 2c shows electrical resistivity as a function of temperature on CaRuO$_3$ ultrathin films grown with different thicknesses between 50 and 35 Å. The 50 Å thick film is metallic, and the electrical resistivity increases as the film thickness decreases. The 35 Å thick film exhibits the maximum electrical resistivity with an upturn in electrical resistivity below 70 K. The electrical resistivity in films less than 26 Å thick decreases again with typical metallic behavior (Fig. 2d). The magnitudes of electrical resistivity systematically increase or decrease as the thin films become thinner. Since the magnitude of the Hall coefficient also corresponds to the electrical resistivity, the electronic state changes with film thickness (Extended Data Fig. 4).

Figure 2e shows plots of electrical resistivity at 4.2 and 300 K on CaRuO$_3$ with different film thicknesses between 8 to 160 Å and a reference thickness of 504 Å. Figure 2f shows the corresponding plots of the sheet conductance to highlight the conductivity. Remarkably, semiconducting and metallic films appear alternately. The maximum resistivity/conductivity values are numbered in ascending order of film thickness and plotted against the film thickness (Fig. 2g). The linearity fitting clarifies that the electrical resistivity oscillates periodically depending on film thickness. The period estimated from the slopes of resistivity and conductivity are 25.1±0.1 and 24.0±1.2 Å, respectively. The 'magic' thickness of 25 Å corresponds to three unit cells of CaRuO$_3$ (110) as shown in Fig. 1a. The resistivity increases below the intercept thickness of ~ 10 Å in the fitting. Since the epitaxiality is confirmed from RHEED observations in the atomic layers (Fig. 1c, Extended Data Fig. 2b), the crystal lattice in CaRuO$_3$ expands thanks to the significant influence of the epitaxial strain from the NdGaO$_3$ (110) substrate[29].

The metal-insulator transitions exhibit crystal anisotropy, and they are not seen or greatly suppressed in the CaRuO$_3$ (001) ultrathin films on a NdGaO$_3$ (001) substrate (Extended Data Fig. 5). The anisotropic electronic state of low dimensional FS is the key to the transitions. The temperature dependence of the electrical resistivity on the 85 Å thick film is well fitted by variable range hopping (VRH) of $\rho \sim \exp[(T_0/T)^{1/(d+1)}]$ with a large excitation energy of $T_0 \sim 26{,}000$ K (Extended Data Fig. 3a). The exponential factor $d = 1$ proves that there is insulation with a strong

Coulomb force. The large cyclotron effective mass observed by the Shubnikov de-Haas effect[26] also indicates that $CaRuO_3$ has a heavy Fermion state. The $CaRuO_3$ is in proximity to the quantum critical point from non-Fermi liquid behavior[18,22-28], which is adjacent to the Mott insulating phase. It can be interpreted that $CaRuO_3$ is similar to the Mott insulator $Ca_2RuO_4$[30].

We perform ultraviolet photoelectron spectroscopy (UPS) on $CaRuO_3$ ultrathin film etched by using argon-ion milling in situ to reproduce the thickness dependence of the properties in one sample. Figure 3a shows the overall spectrum of the $CaRuO_3$ ultrathin film etched every a second from 0 to 15 seconds. The initial film thickness is 110 Å. The peak derived from the oxygen state seen around the binding energy (BE) = 5.6 eV is high in the non-etched film[31]. Considering the probe depth of a few angstrom realized with He I phonons, the calcium ions on the surface of the film form strong bonds with oxygen. The valence band represents an insulating layer with low conduction. The oxygen peak is sharpened after 1 second of etching. The noticeable increase in the valence band below 3 eV exhibits the $CaRuO_3$ spectrum. After 7 seconds of etching, the optical intensity reaches its maximum value. It starts to decrease after 8 seconds, and no change is observed after 11 seconds, which indicates that the $CaRuO_3$ is completely removed. Figure 3b shows a close-up spectrum of another $CaRuO_3$ film with BE between 0.5 and 3.5 eV to measure in detail the result of every half second etching. The intensity tends to increase by etching with good reproducibility compared with the previous sample. A crossover can be observed between 0.9 and 1.6 eV after 1.0 - 2.0 seconds of etching. The hump was reported to originate from Ru-4$d$ electrons as observed by angle resolved photoelectron spectroscopy[18,32]. It was mentioned that phonons, polarons, or magnons could not contribute to the hump because of the large energy[32], which can be confirmed from the large $T_0$.

Figure 3c shows the UPS with BE between 0 and 3.5 eV as a function of etching time. The intensity tends to increase as the etching progresses below 3 eV. It is easy to increase the optical intensity by ion milling, which reveals that the surface roughness causes the destruction of the insulation. A stepped structure can be seen, especially between 1.0 and 2.0 eV, which is caused by the hump. Since the ultrathin film is etched from 110 Å to 0 Å for 11 seconds, it is comparable to the film thickness dependence shown in Fig. 2. Figure 3d compares the sheet conductance as a function of thickness at 300 K and the optical intensity of UPS at 1.1 eV as a function of etching time. The arrows represent the coincidence between the increase in optical intensity and the peak of the electrical conduction at etching time $t$ = 1.0, 4.0, and 6.0 sec. Finally, it is reproducibly proven on one etched film that the periodicity of the thickness in optical intensity is the same as that in electrical conductivity. The periodicities show good agreement at BE = 1.1 eV where the hump is prominent, which reveals that suppression on the Ru-4$d$ orbital is related to insulation.

What is the mechanism of the metal-insulator transitions that oscillate with thickness? The important features are i) the transitions alternate with the magic thickness period of 25 Å corresponding to three times the thickness of a CaRuO$_3$ (110) unit cell, ii) the transitions are accompanied by a large excitation energy equivalent to a few electron volts, iii) anisotropy, namely they can be clearly seen only in CaRuO$_3$ (110) growth. Until now, it has been proposed that quantum size effects are the oscillation of properties depending on scale[1-3]. However, the energy that accompanies metal-insulator transitions is too large to be interpreted as quantum wells within the bandwidth as small as ~ 200 meV in CaRuO$_3$[18]. When the strong Coulomb interaction in quantum critical phase is taken into consideration, it is necessary to discuss the Mott insulation for a large excitation energy.

Although typical Mott insulators appear in half filling bands, compounds that have no half filling bands also become Mott insulators as a result of the stripe correlations of spins and carriers[33]. A concept of commensurability on stripe Mott insulation is suitable for 2D-systems with several unit cells. It is considered that the Mott insulation on CaRuO$_3$ (110) ultrathin films is dependent on the commensurability on stripe correlations when the thickness is a multiple of three unit cells (Fig. 4a, c). Considering the period of 25 Å for the oscillating transitions, it is assumed that there is the single RuO$_6$ layer of domain wall for every three unit cells. The other RuO$_6$ layers form the Mott insulator with anti-ferromagnetism. Since electrons are left over except for the multiple of the magic thickness, the Mott insulating stripes on CaRuO$_3$ is dissolved and reverts to metal (Fig. 4b).

The stripe correlations of spins and carriers with three time periods are triggered by Peierls instability[34-38]. CaRuO$_3$ consists of a square pole FS[18], which is advantageous for nesting properties. Since the wavenumber $k_{[110]}$ of the FS corresponds to one-third of the Brillouin zone, the energy gain from the stabilization of Peierls transition is large enough to form the period of 25 Å in real space. The Peierls instability is proven by the anisotropy of the metal-insulator transitions. They are clearly observed on CaRuO$_3$ (110) where the nesting property is excellent in the growth direction. By contrast, they can be hardly seen on CaRuO$_3$ (001) films where the pole FS has an open orbit in the growth direction [001], i.e., it cannot take any nesting vectors. Namely, the metal-insulator transitions are caused by the dimensional restrictions in different directions caused by each 2D-FS and the ultrathin films. The metal-insulator transitions are kinds of as the incommensurability-commensurability transitions that occur only under clear boundary conditions. No scaling dependences on CaRuO$_3$ have been discovered in thick films or bulk samples because of the mean free path of electrons.

We concluded that the metal-insulator transitions oscillating with thickness originates from the commensurability of Mott insulating stripes assisted by Peierls instability arising from the dual restriction on dimensions in wavenumber space and real space. We establish 'scales' as a new parameter for quantum phase transitions following the angle[39]. We anticipate that novel quantum properties caused by commensurability will appear in nano-scaling samples of electron correlated compounds.

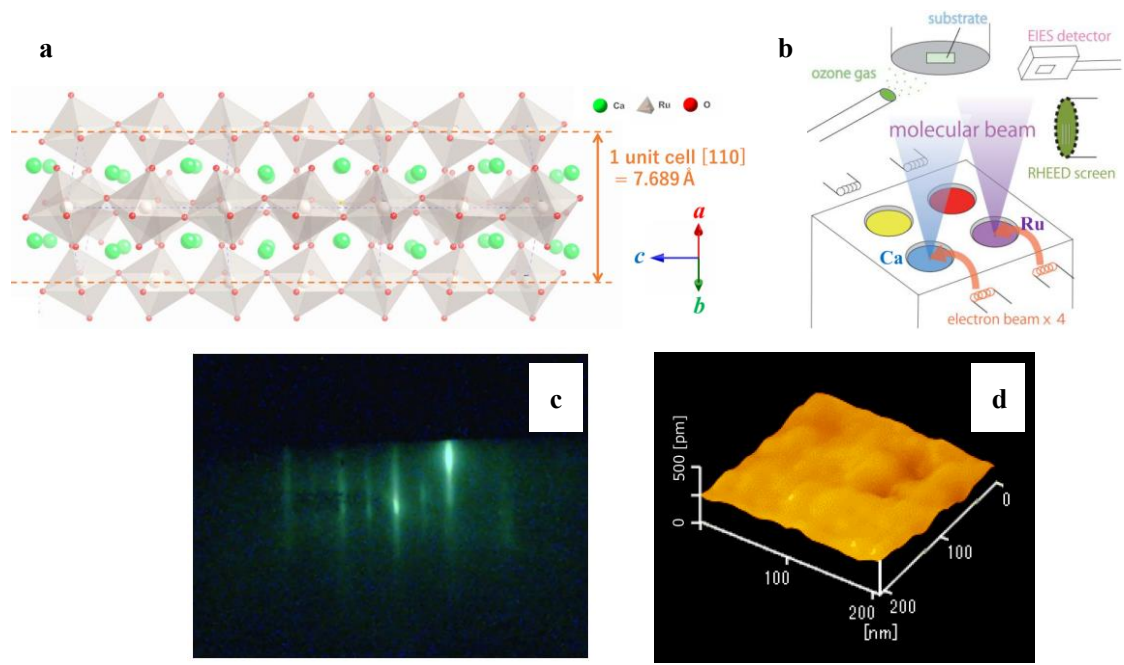

**Fig. 1 | Preparation. a,** Crystal structure of $CaRuO_3$ grown on $NdGaO_3$ (110) substrate. **b,** Schematic diagram of molecular beam epitaxy method. **c,** RHEED pattern of 8 Å thick $CaRuO_3$ films. **d,** Roughness of surface on 50 Å thick $CaRuO_3$ ultrathin film observed with AFM.

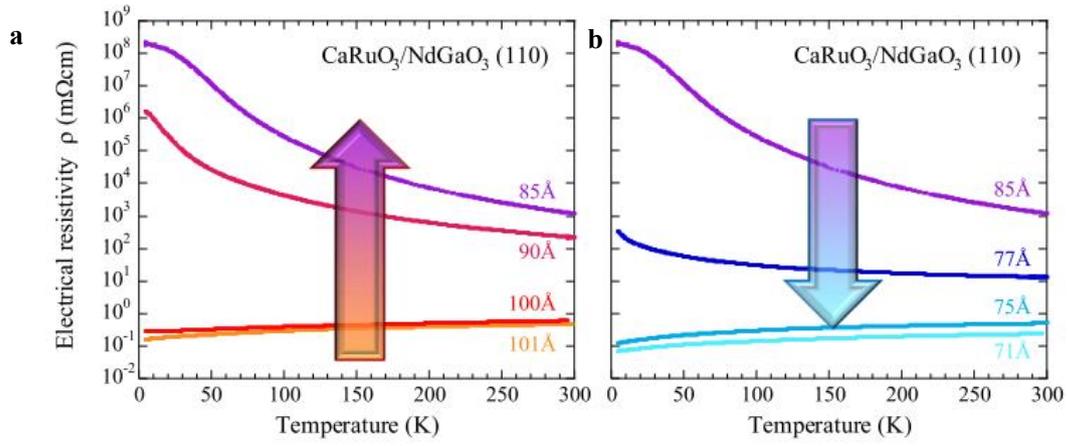
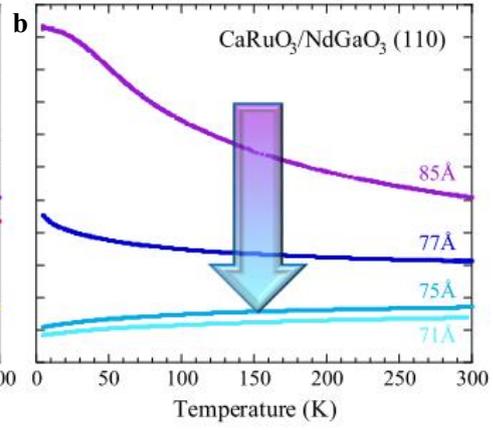
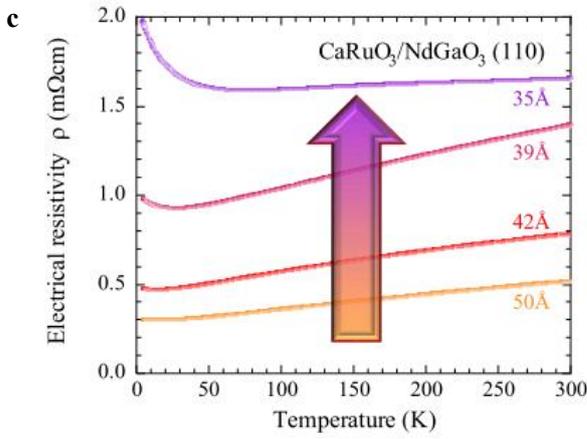
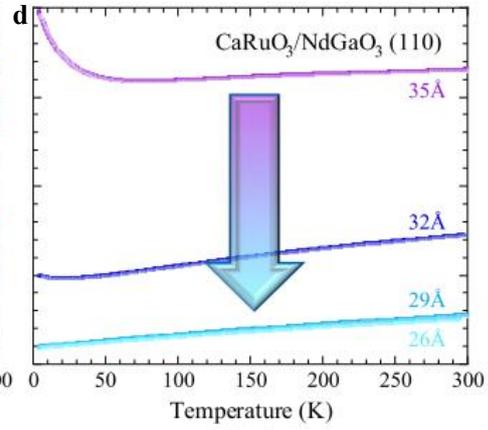
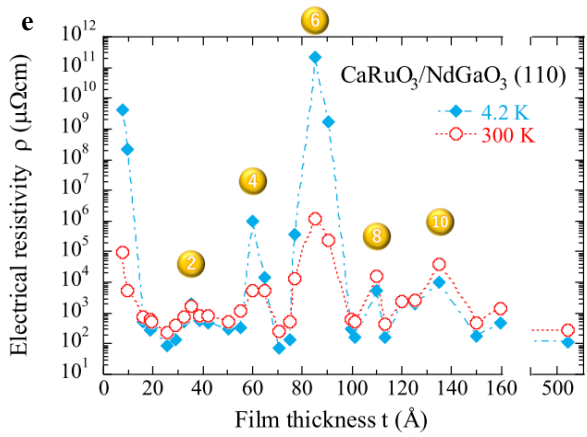
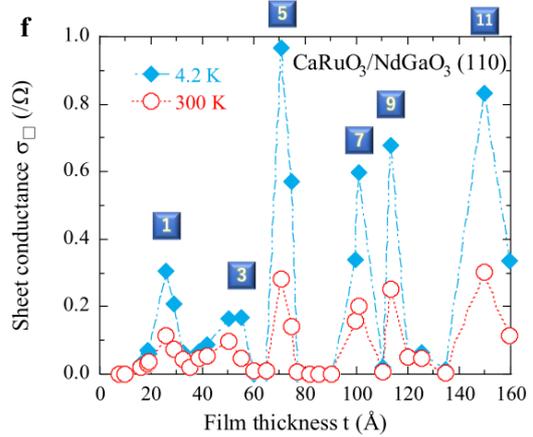

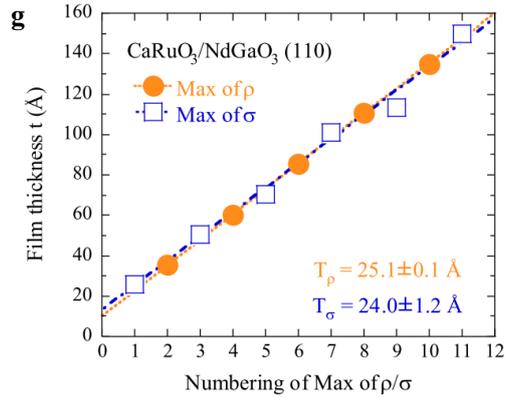

**Fig. 2 | Thickness dependence of electrical resistivity on CaRuO₃ ultrathin films. a,b,** Electrical resistivity on CaRuO₃ thin films with thicknesses between 101 to 71 Å as a function of temperature. **c,d,** Electrical resistivity on CaRuO₃ thin films with thicknesses between 50 to 26 Å as a function of temperature. **e,f,** Electrical resistivity and sheet conductance as a function of thickness on CaRuO₃ films at 4.2 and 300 K, respectively. **g,** Periodicity of film thickness on maximum of resistivity/conductance at low temperature.

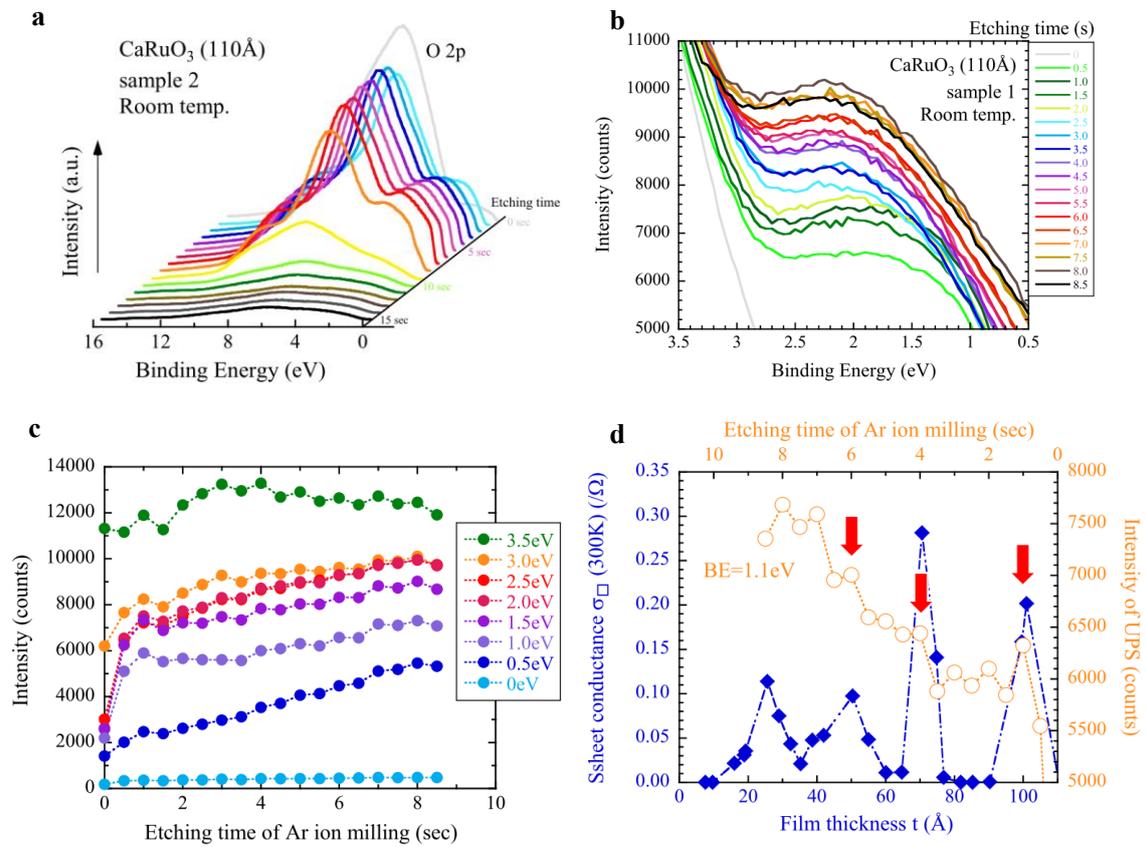

**Fig. 3 | Correspondence of periodic transitions in optical intensity to a CaRuO$_3$ ultrathin film etched by argon ion milling. a,** Overall UPS spectra of 110 Å thick CaRuO$_3$ ultrathin film etched from 0 to 15 seconds in situ. **b,** Details of spectrum with BE between 0.5 and 3.5 eV of the other 110 Å thick CaRuO$_3$ ultrathin film etched every 0.5 seconds from 0 to 8.5 seconds. **c,** Optical intensity as a function of etching time with BE from 0 to 3.5 eV. **d,** Comparison of electrical conductance as a function of film thickness and optical intensity as a function of etching time. The arrows indicate the correspondence to the periodicity of the oscillating transitions.

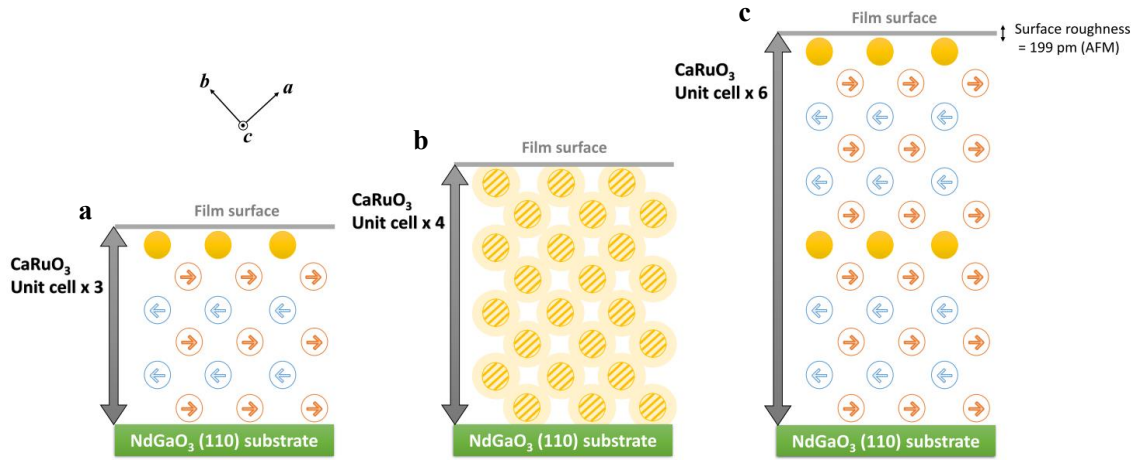

**Fig. 4 | Commensurability schematic of Mott insulation on CaRuO₃ (110) ultrathin films.** The film thicknesses are **a,** 3 units, **b**, 4 units, and **c,** 6 units. The circles indicate the Ru ions. The closed circles indicate the domain walls occupied by electrons or holes. The open circles indicate the Mott insulating localized electrons. The arrows indicate the anti-ferromagnetism state. The circles with diagonal lines indicate the metallic state of non-filled bands.

## Methods

### Film fabrication

Ultrathin films are prepared in a customer-designed MBE chamber (basal pressure of ~ 5 x $10^{-6}$ Pa). Molecular beams of Ru and Ca are controlled by electron impact emission spectroscopy (EIES). The molecular beam of $CaF_2$ is detected with a quartz crystal microbalance method in a constant power mode. The target rates of the molecular beam are Ru 0.1 Å/s, Ca 0.3 Å/s $CaF_2$ ~ 1 Å/s, respectively. The films are totally grown at rate of 25 Å/min. We determine the film thickness from the deposition time. The oxygen is supplied by $O_3$ using a conventional ozonizer (The ozone concentration is about 14 %). The pressure of the $O_2$ in the chamber is monitored with a quadrupole mass electrometer (Q-mass). The growth condition is ~2 x $10^{-4}$ Torr for all the films. The growth temperature is 800 °C. The home-made MBE system is detailed elsewhere[16]. The substrate is $NdGaO_3$ (110) with a typical size of ~3 mm x 6 mm. The surface roughness of the $NdGaO_3$ substrate is typically below 2.5 Å.

The $CaRuO_3$ films grow along the same crystal orientation as orthorhombic $NdGaO_3$ substrates because the lattice constant is very close to that of $NdGaO_3$[19,20]. The high-quality epitaxial films are grown by the MBE method because the excess ruthenium supplied re-evaporates as $RuO_x$ in an oxygen atmosphere[15]. Since calcium has a high vapor pressure, the excess calcium supplied also re-evaporates from the substrate surface. Accordingly, both Ca and Ru are autoregulating constituent elements on epitaxial growth. We reproducibly fabricate pure and highly crystalline $CaRuO_3$ ultrathin films using a home-made MBE system[16] (Fig. 1b). $CaRuO_3$ (110) is grown on a $NdGaO_3$ (110) substrate with crystal periodicity of 7.689 Å and containing two layers of $RuO_6$ octahedra[12,13] (Fig. 1a).

We deposit 50 Å of $CaF_2$ insulator as a barrier layer on $CaRuO_3$ ultrathin films below 200 °C as shown in Extended Data Fig. 1a. It is difficult to measure the electrical resistivity of atomic layer films reproducibly without the barrier due to the high resistivity, which destroys the film surface in the atmosphere. The barriered films have less surface roughness because they are not exposed to the atmosphere. We observe the surface of the films in-situ using RHEED. The growth direction of $CaRuO_3$ (110) on the $NdGaO_3$ (110) substrate is confirmed by electron back-scattering diffraction (EBSD) as shown in Extended Data Fig. 1e.

### Measurements

We employ the $CaRuO_3$ ultrathin film to measure the X-ray diffraction (XRD) shown in Extended Data Fig. 1c, d. We also observe the surfaces of the films using field emission-scanning electron microscopy (FE-SEM) (JSM-7000F, JEOL) and an atomic force microscope (AFM) (SPA-400). We measure the electrical properties in helium by using the conventional four-probe and two-probe methods for metal and insulating samples, respectively. The electrode is soldered with indium to destroy the $CaF_2$ barrier layer as shown in Extended data Fig. 1b. Some samples are measured at temperatures of as low as 0.5 K using a home-made $^3$He refrigerator. We also measure the electrical resistivity and Hall effect at temperatures as low as 2.2 K with a magnetic field of up to 6 T by using a physical property measurement system (PPMS, Quantum Design). The measurement setup is shown in Extended data Fig. 1b where a 6-probe method is used to measure electrical resistivity and Hall effects. The UPS measurements and argon ion milling are performed with a JPS-9010MC (JEOL). The data is collected at room temperature with He Iα ($h\nu$ = 21.2 eV) photons. The energy resolution is set at 50 meV.

**Acknowledgements**

The authors thank Prof. M. Naito, K. Ichimura, R. Suzuki, N. Matsunaga, T. Kurosawa, Dr. Y. B. Nanao, and K. Nakatsugawa for valuable discussions and support. A part of this work was conducted at the Laboratory of XPS Analysis, Joint-use Facilities, Hokkaido University. The Open Facility of the Institute for Catalysis was used for the UPS. This study is financially supported by Takahashi Industrial and Economic Research Foundation, The Japan Prize Foundation, The Murata Science Foundation, The Yashima Environment Technology Foundation, The Inamori Foundation, The Iketani Science and Technology Foundation, and The Samco Science and Technology Foundation.

**Author Contributions**

M. S. planned the research. M. S. grew the ultrathin films. M. S. performed the RHEED, AFM, XRD, and EBSD observations. M. S., and H. N. measured the electrical resistivity, and the Hall effects. M. S. and S. S. measured PES. M. S. wrote the manuscript. M. S., H. N, S. S., and S. T. discussed the results and commented on the manuscript. S. T. supervised the project.

**Competing interests**

The authors declare no competing interests.

**Materials & Correspondence**

Masahito Sakoda, Ph.D.

Department of Applied Physics, Graduate School of Engineering, Hokkaido University

Kita 13-Nishi 8 chome, Kita-ku, Sapporo, Hokkaido 060-8628, Japan

Tel.　+81-11-706-6154

E-mail　sakodam@eng.hokudai.ac.jp

**Extended Data Figures**

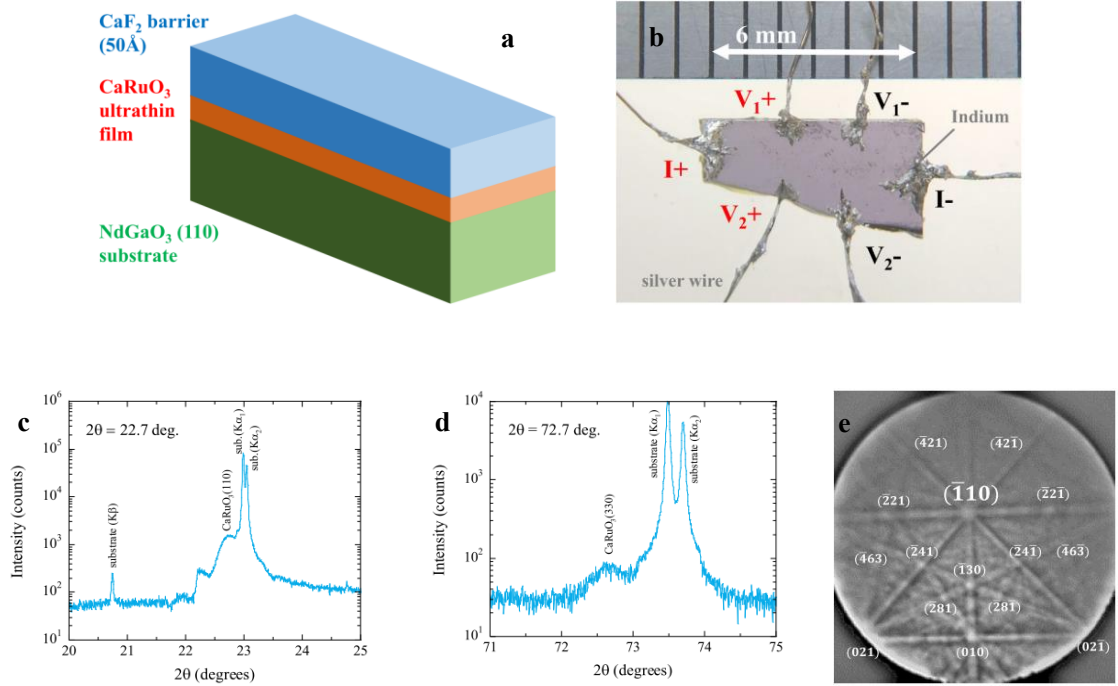

**Extended Data Fig. 1 | Sample preparation. a,** Schematic of a CaRuO$_3$ ultrathin film covered by a CaF$_2$ barrier layer. **b,** Picture of a typical sample of CaRuO$_3$ ultrathin film on a NdGaO$_3$ (110) substrate. The substrates are as long as several millimeters. $\theta$-$2\theta$ scans on X-ray diffraction around peaks of **c,** (110) and **d,** (330) on CaRuO$_3$ thin film with a thickness of 170 Å. **e,** EBSD pattern of 50 Å thick CaRuO$_3$ film.

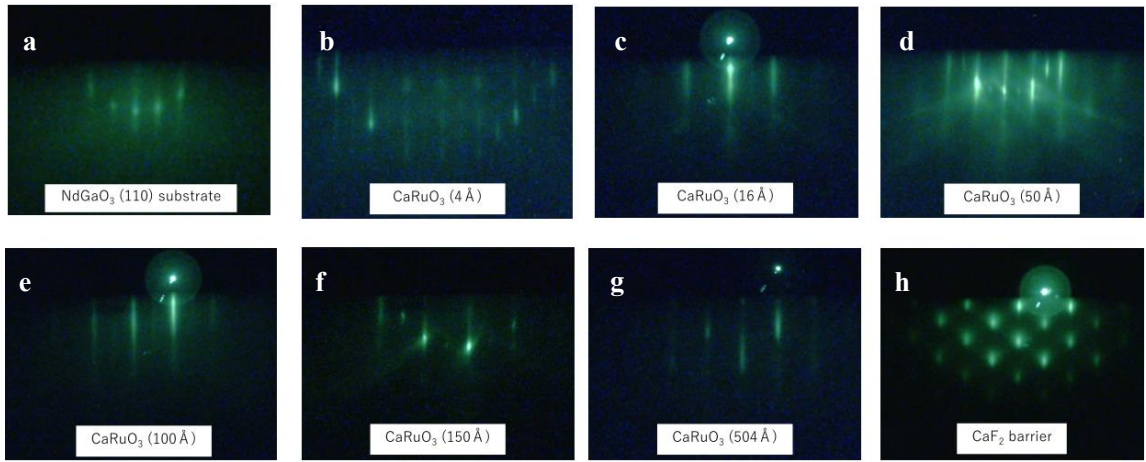

**Extended Data Fig. 2 | RHEED patterns on CaRuO₃ ultrathin film for characteristic thickness, reference substrate and CaF₂ barrier. a,** Typical pattern of a NdGaO₃ (110) substrate is streaky. **b, c, d, e, f, g,** Patterns of CaRuO₃ films with thicknesses of 4, 16, 50, 100, 150, 504 Å, respectively. They exhibit sharp streaks, especially on the films with thicknesses of 8 – 150 Å. **h,** The typical pattern of a CaF₂ barrier is spotty.

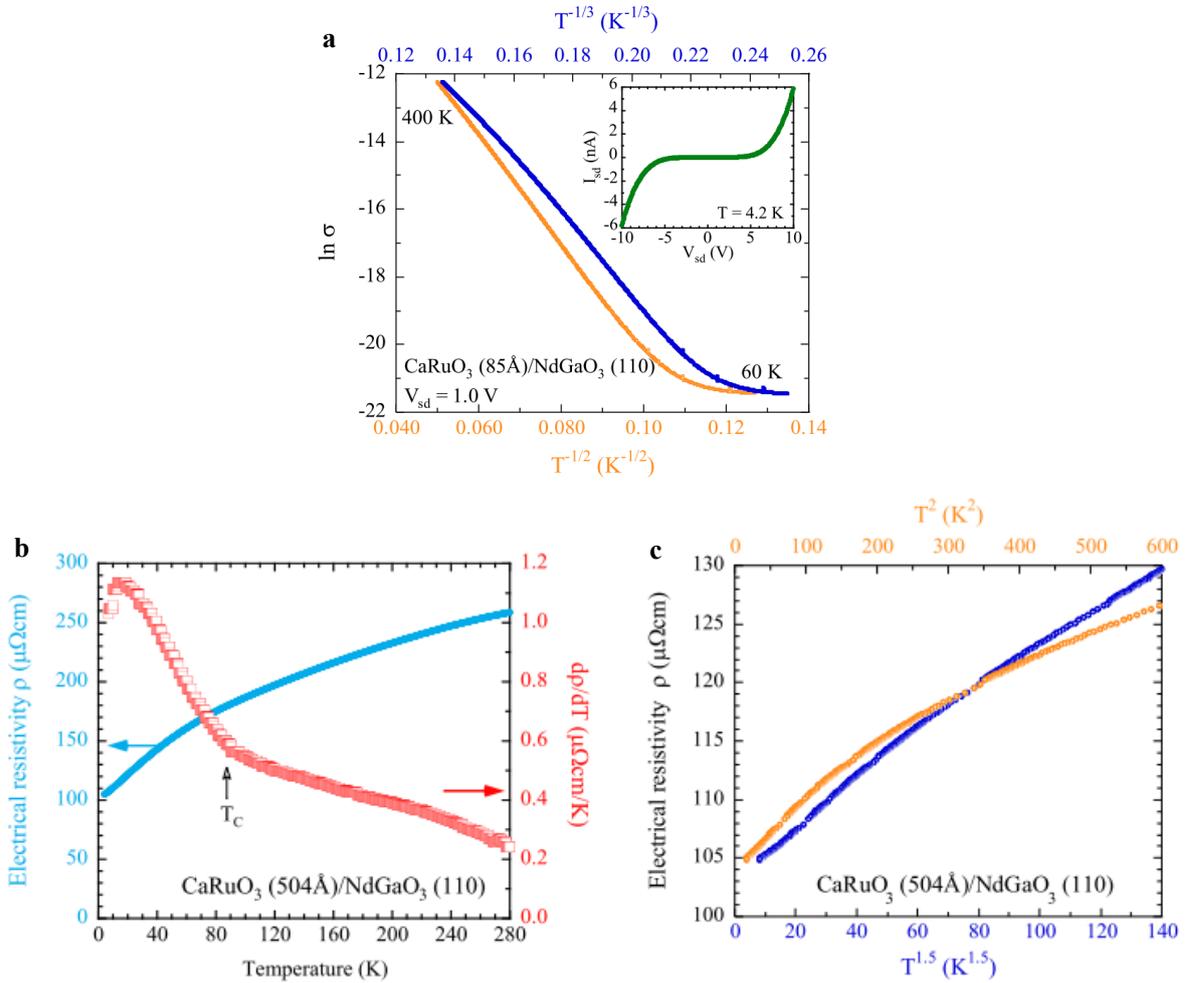

**Extended Data Fig. 3 | Analysis of temperature dependence of electrical resistivity. a,** Logarithm of electrical conductance as a function of $T^{1/2}$ (lower), and $T^{1/3}$ (upper) on 85 Å thick CaRuO$_3$ ultrathin film. The inset shows the current-voltage curve at 4.2 K. **b,** Temperature dependence of electrical resistivity ρ (left) and differential resistivity dρ/dT (right) of CaRuO$_3$ film with a reference thickness of 504 Å. **c,** Plots of electrical resistivity as a function of $T^{1.5}$ (lower) and $T^2$ (upper) at low temperature.

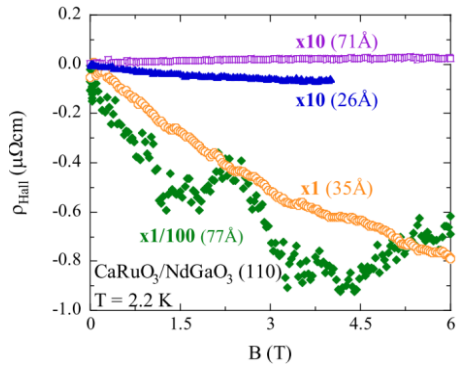 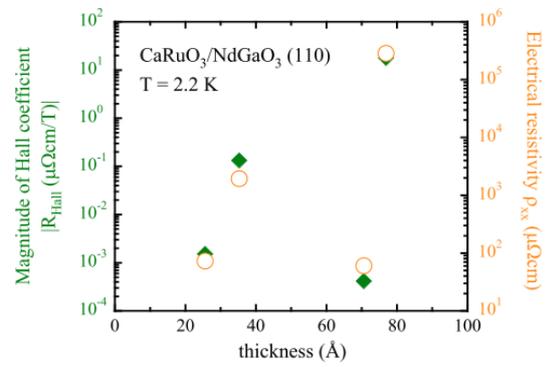

**Extended Data Fig. 4 | Hall coefficients of CaRuO₃ ultrathin films. a,** Comparison of Hall coefficients as a function of magnetic flux density $B$ on CaRuO$_3$ ultrathin films with different thicknesses of 26, 35, 71, 77 Å. **b,** Comparison of the magnitude of Hall coefficients (left) and electrical resistivity (right) at 2.2 K as a function of film thickness.

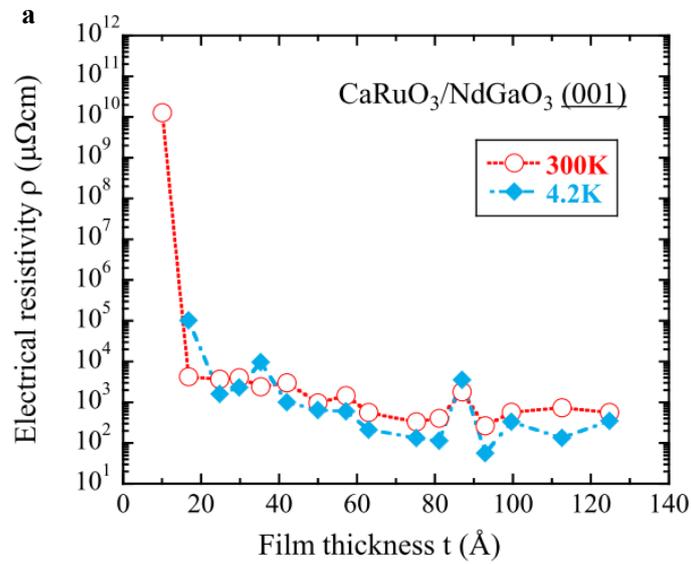
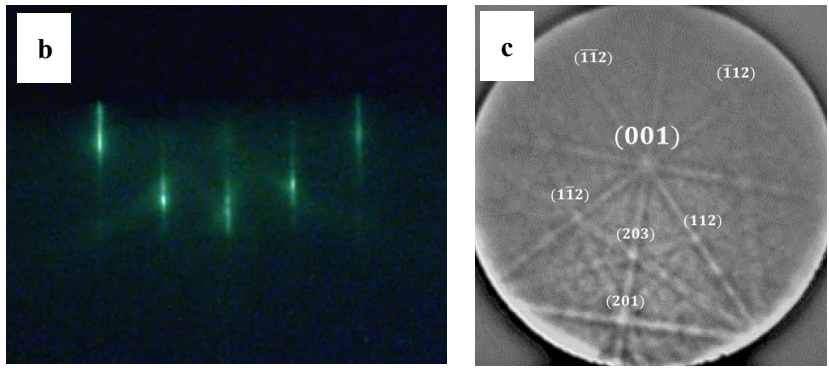

**Extended Data Fig. 5 | Properties of CaRuO₃ (001) grown on NdGaO₃ (001) substrate. a,** Electrical resistivity as a function of thickness at 300 K and 4.2 K. **b,** RHEED pattern of CaRuO₃ (001) with thickness of 25 Å. It shows a sharp streak pattern, which means a flat surface and high crystallinity. **c,** EBSD pattern of 500 Å thick CaRuO₃ (001) film. CaRuO₃ grows in the [001] crystal direction on a NdGaO₃ (001) substrate.